\documentclass[preprint,aps,prl,fleqn,showpacs]{revtex4}

\usepackage{color} 

\usepackage{graphicx}
\usepackage{bm}
\usepackage{amsmath}
\usepackage[dvipdfm,bookmarks=true,bookmarksnumbered=true,bookmarkstype=toc]{hyperref} 

\begin{document}
\newcommand{\Area}{A}
\newcommand{\Av}{{\bm A}}
\newcommand{\av}{{\bm a}}
\newcommand{\Bv}{{\bm B}}
\newcommand{\DOS}{{\nu}}
\newcommand{\dx}{{d^3 x}}
\newcommand{\eF}{{\epsilon_F}}
\newcommand{\etaz}{{\eta^0}}
\newcommand{\etazk}{\eta^0_k}
\newcommand{\etazmk}{\eta^0_{-k}}
\newcommand{\Ev}{{\bm E}}
\newcommand{\js}{{j_{\rm s}}}
\newcommand{\jsc}{{j_{\rm s}^{\rm cr}}}
\newcommand{\jsca}{{j_{\rm s}^{\rm cr (1)}}}
\newcommand{\jscb}{{j_{\rm s}^{\rm cr (2)}}}
\newcommand{\jc}{{j^{\rm cr}}}
\newcommand{\kF}{{k_F}}
\newcommand{\kp}{{k_\perp}}
\newcommand{\Kp}{{K_\perp}}
\newcommand{\kB}{{k_B}}
\newcommand{\kBT}{{k_B T}}
\newcommand{\kv}{{\bm k}}
\newcommand{\Mv}{{\bm M}}
\newcommand{\nv}{{\bm n}}
\newcommand{\pv}{{\bm p}}
\newcommand{\Ne}{{N_{\rm e}}}
\newcommand{\Omegazk}{{\Omega^{(0)}_{\kv}}}
\newcommand{\Omegaok}{{\Omega^{(1)}_{\kv}}}
\newcommand{\omegazk}{{\omega^{(0)}_{\kv}}}
\newcommand{\Omegak}{{\Omega_{\kv}}}
\newcommand{\omegaell}{{\omega_{\ell}}}
\newcommand{\phiz}{{\phi_0}}
\newcommand{\phizc}{\bar{\phi_0}}
\newcommand{\qv}{{\bm q}}
\newcommand{\rv}{{\bm r}}
\newcommand{\rhow}{{\rho_{\rm w}}}
\newcommand{\sigmav}{{\bm \sigma}}
\newcommand{\Sv}{{\bm S}}
\newcommand{\Smh}{{\left(S-\frac{1}{2}\right)}}
\newcommand{\Tv}{{\bm T}}
\newcommand{\torq}{{\tau}}
\newcommand{\Tz}{{\tau}}
\newcommand{\tileta}{{\tilde\eta}}
\newcommand{\tilF}{\tilde{F}}
\newcommand{\tilFc}{\tilde{F}_{\rm c}}
\newcommand{\tilK}{{\tilde{K}}}
\newcommand{\tilKp}{{\tilde{K}_\perp}} 
\newcommand{\tiln}{{\tilde{n}}}
\newcommand{\tilTz}{v_{\torq}}
\newcommand{\tilv}{{\tilde{v}}}
\newcommand{\thetaz}{{\theta_0}}
\newcommand{\Vz}{{V_0}}
\newcommand{\xv}{{\bm x}}
\newcommand{\xiz}{\xi_0}
\newcommand{\Vv}{{\bm V}}
\newcommand{\vv}{{\bm v}}
\newcommand{\vc}{v_{\rm c}}
\newcommand{\vck}{v_{\rm c}^{K}}
\newcommand{\vci}{v_{\rm ins}}
\newcommand{\beqa}{\begin{eqnarray}}
\newcommand{\eeqa}{\end{eqnarray}}
\newcommand{\lrp}{\stackrel{\leftrightarrow}{\partial}}
\newcommand{\vef}{\varepsilon_{\rm F}}
\newcommand{\vare}{\varepsilon}

\title{Inverse Spin Hall Effect Driven by Spin Motive Force}
\author{Junya Shibata}
\email{shibata@gen.kanagawa-it.ac.jp}
\affiliation{Kanagawa Institute of Technology, 
1030 Shimo-Ogino Atsugi, Kanagawa 243-0292, Japan}
\author{Hiroshi Kohno}
\affiliation{
Graduate School of Engineering Science, Osaka University,
Toyonaka, Osaka 560-8531, Japan}%
\date{\today}

\begin{abstract}
The spin Hall effect is a phenomenon that an electric field 
induces a spin Hall current. 
In this Letter, we examine the inverse effect that, 
in a ferromagnetic conductor, 
a charge Hall current is induced by a spin motive force, or a spin-dependent 
effective \lq electric' field ${\bm E}_{\rm s}$, arising from the 
time variation of magnetization texture. 
By considering skew-scattering and side-jump processes 
due to spin-orbit interaction at impurities, we obtain 
the Hall current density as $\sigma_{\rm SH} {\bm n}\times{\bm E}_{\rm s}$, 
where ${\bm n}$ is the local spin direction and $\sigma_{\rm SH}$ 
is the spin Hall conductivity. 
The Hall angle due to the spin motive force 
is enhanced by a factor of $P^{-2}$ 
compared to the conventional anomalous Hall effect 
due to the ordinary electric field, 
where $P$ is the spin polarization of the current. 
The Hall voltage is estimated for a field-driven domain wall 
oscillation in a ferromagnetic nanowire. 
\end{abstract}
\maketitle

{\it Introduction:} 
 Magnetization dynamics induced by an electric current flowing 
in a nano-structured ferromagnet 
has been studied intensively for a decade 
because of the enormous application potentialities 
called spintronics. 
 It has been well recognized that such phenomena 
are due to spin torques 
\cite{{Berger92},{Slonczewski96}} that 
localized spins of $d$-electrons in a ferromagnet 
are exerted by conducting $s$-electrons through 
the $s$-$d$ exchange coupling.

It was proposed that as a reaction to spin torques there arises a 
spin-dependent motive force (spin motive force) 
from magnetization dynamics 
\cite{{Berger86},{Stern92},{BM07},{Saslow07},{Duine08},
{Tserkovnyak08},{YXN07},{YBKXNTE08}}. 
 For a slowly-varying spin texture ${\bm n}$ 
(and in the absence of spin relaxation), 
it is expressed by the spin-dependent effective \lq electric' field as 
\cite{{Duine08},{Tserkovnyak08}}
\beqa
 E_{{\rm s},i} = \frac{\hbar}{2e}\nv\cdot(\partial_{i}\nv\times\dot{\nv}) .
\label{Es}
\eeqa  
The field ${\bm E}_{\rm s}$, or the force 
${\bm F}_{\rm s} = -e{\bm E}_{\rm s}$, 
acts on the electrons in a spin-dependent way, namely, it 
drives majority-spin and minority-spin electrons in mutually opposite 
directions \cite{spin_dependence} and produces a (diagonal) spin current in the direction of 
${\bm E}_{\rm s}$. 
In the presence of spin-orbit interaction (SOI), 
the orbits of opposite-spin electrons 
will be curved in opposite directions, 
and a net Hall current is expected 
in a direction perpendicular to ${\bm E}_{\rm s}$. 

Similar phenomenon was proposed as the inverse spin Hall effect (ISHE) 
where a spin current is converted 
to a charge current via SOI, and observed experimentally 
\cite{{Saitoh06},{Tinkham06},{Kimura07},{ATHSIMS08},{Seki08}}.
 Theoretical studies were given for nonmagnetic metals with a spin 
current injected from the attached ferromagnet by the spin-pumping effect 
due to spin dynamics \cite{{WBWBT06},{OTT07},{TT08},{XBB08}}.

In this Letter, 
we study the ISHE induced by spin motive force, or ${\bm E}_{\rm s}$, 
due to the dynamics of spin texture in ferromagnetic metals, 
including SOI from impurities. 
We will show that the total current is given by 
\beqa
\label{total_current}
\mathcal{J} = \sigma_{\rm s}{\bm E}_{\rm s} 
 + \sigma^{\phantom{\dagger}}_{\rm SH}
\nv\times{\bm E}_{\rm s}, 
\eeqa
where $\sigma_{\rm s} = \sigma_{\uparrow}-\sigma_{\downarrow}$ 
is the \lq\lq spin conductivity" and 
$\sigma^{\phantom{\dagger}}_{\rm SH} 
= \sigma^{\phantom{\dagger}}_{{\rm H} \uparrow} 
+ \sigma^{\phantom{\dagger}}_{{\rm H} \downarrow}$ 
is the spin Hall conductivity, 
with $\sigma_{\uparrow}$ and $\sigma^{\phantom{\dagger}}_{{\rm H} \uparrow}$ 
($\sigma_{\downarrow}$ and $\sigma^{\phantom{\dagger}}_{{\rm H} \downarrow}$)
being diagonal and Hall conductivities for majority-spin (minority-spin) 
electrons. 

 Equation (\ref{total_current}) may be contrasted with two 
related phenomenon in ferromagnets. 
 One is the spin Hall effect \cite{Hirsch99}, 
given by the second term of the relation 
\beqa
\label{total_spin_current_E}
\mathcal{J}_{\rm S} = \sigma_{\rm s}{\bm E} 
+ \sigma^{\phantom{\dagger}}_{\rm SH}
\nv\times{\bm E}, 
\eeqa
which shows that a spin current 
$\mathcal{J}_{\rm S} = \mathcal{J}_\uparrow - \mathcal{J}_\downarrow$ 
is induced by an ordinary electric field, ${\bm E}$. 
The other is the anomalous Hall effect \cite{DCB01}, 
\beqa
\label{total_current_E}
\mathcal{J} = \sigma_{\rm c}{\bm E} 
+ \sigma^{\phantom{\dagger}}_{\rm H}
\nv\times{\bm E}, 
\eeqa
where $\sigma_{\rm c} = \sigma_{\uparrow} + \sigma_{\downarrow}$ 
is the electrical (\lq\lq charge") conductivity and 
$\sigma^{\phantom{\dagger}}_{\rm H} 
= \sigma^{\phantom{\dagger}}_{{\rm H} \uparrow} 
- \sigma^{\phantom{\dagger}}_{{\rm H} \downarrow}$
is the anomalous Hall conductivity.

 The Hall resistivity, 
$\rho_{\rm SH} = \sigma_{\rm SH}/\sigma_{\rm s}^2$, 
in the present case Eq. (\ref{total_current}) 
is larger by a factor of $\sim P^{-3}$ compared to that of the 
conventional AHE, $\rho_{\rm H} = \sigma_{\rm H}/\sigma_{\rm c}^2$, 
where 
$P = \sigma_{\rm s}/\sigma_{\rm c}$ 
$(\simeq \mathcal{J}_{\rm S}/\mathcal{J})$ 
is the spin polarization of the current. 
 The result will be applied to an oscillating motion of a domain wall 
driven by a magnetic field, and the Hall voltage is estimated.

 Usually, the relation Eq. (\ref{total_current_E}) assumes a uniform 
magnetization, ${\bm n} = \hat z$ for example. 
 The derivation of Eq. (\ref{total_current}) presented in this Letter 
also justifies  
Eqs. (\ref{total_spin_current_E}) and (\ref{total_current_E}) 
generalized to the case of slowly-varying ${\bm n}$.

{\it Model:} 
 We consider a ferromagnetic metal containing impurities with SOI. 
 We adopt the {\it s-d} model consisting of conduction {\it s}-electrons 
and localized {\it d}-electron spins, 
both are coupled ferromagnetically. 
 The localized $d$-spins are treated as classical, 
and assumed to be slowly varying in space and time. 
 They are denoted by ${\bm S}(\rv,t) = S\nv(\rv,t)$, 
where $S$ is the magnitude of the $d$-spin and 
$\nv=(\sin\theta\cos\phi,\sin\theta\sin\phi,\cos\theta)$ 
is a unit vector. 
 The total Lagrangian of the $s$-electron system is given by 
$L=L_{0}-H_{\rm sd}-H_{\rm so}$, 
\beqa
&& \hskip -5mm 
 L_{0} = \int d^3\rv~c^{\dagger}\left(
i\hbar\frac{\partial}{\partial t}
+\frac{\hbar^2}{2m}\nabla^{2}
+\vare_{\rm F} -V_{\rm imp}\right)c, \\
&& \hskip -5mm 
 H_{\rm sd}=-M\int d^{3}\rv~\nv(x)\cdot(c^{\dagger}{\bm \sigma}c)_{x}, \\
&& \hskip -5mm 
 H_{\rm so}= \lambda_{\rm so} \frac{m}{\hbar}\vare_{ij\alpha}
\int d\rv~(\partial_{i}V_{\rm imp}(\rv))j_{j}^{\alpha}(x), 
\label{Lagrangian_so}
\eeqa
where $c^{\dagger}(x)=(c^{\dagger}_{\uparrow}(x),
c^{\dagger}_{\downarrow}(x))$ 
is the electron creation operator at $x=(\rv,t)$, 
$\vare_{\rm F}$ is the Fermi energy, 
$2M$ is the $s$-$d$ exchange splitting, and 
$\bm \sigma$ is a vector of Pauli spin matrices. 
 The impurity potential is modeled as the short-ranged one, 
$V_{\rm imp}(\rv) = u\sum_{i}\delta(\rv - {\bm R}_{i})$, 
where $u$ denotes the strength of the impurity potential and 
${\bm R}_{i}$ represents the randomly distributed impurity positions. 
 The $H_{\rm so}$ describes SOI at impurities, 
where 
$j^{\alpha}_{j}=\displaystyle\frac{\hbar}{2mi}
(c^{\dagger}\sigma^{\alpha}
\overset{\leftrightarrow}{\partial}_{j}c)
=\displaystyle\frac{\hbar}{2mi}
(c^{\dagger}\sigma^{\alpha}\partial_{j}c
-(\partial_{j}c^{\dagger})\sigma^{\alpha}c)$ 
is the spin-current density, 
$\lambda_{\rm so}$ is the strength of SOI,  
and 
$\vare_{i j \alpha}$ is the complete anti-symmetric tensor 
with $\varepsilon_{xyz}=1$. 
Repeated index implies summation over 
$i,j, \alpha = x,y,z$.

In ferromagnetic metals, 
the exchange coupling energy $M$ is strong, and 
it is useful to perform a local transformation 
so that the spin quantization axis of $s$-electrons is taken to be the 
local $d$-spin direction $\nv$ at each point 
of space and time \cite{{KMP77},{Volovik87},{TF94}}; 
$c(x) = U(x)a(x)$, $c^{\dagger}(x)= a^{\dagger}(x)U^{\dagger}(x)$, 
where $U$ is a 2 $\times$ 2 unitary matrix given by 
$U={\bm m}\cdot{\bm \sigma}$ with ${\bm m}=(\sin(\theta/2)\cos\phi, 
\sin (\theta/2)\sin\phi, \cos(\theta/2))$. 
The spin density $c^{\dagger}\sigma^{\alpha}c$ is 
transformed into $a^{\dagger}U^{\dagger}\sigma^{\alpha}Ua 
={\cal R}^{\alpha\beta} a^{\dagger}\sigma^{\beta}a$, 
where ${\cal R}^{\alpha\beta}=2m^{\alpha}m^{\beta}-\delta^{\alpha\beta}$ 
is a $3\times 3$ orthogonal matrix. 
Noting that ${\cal R}^{\alpha\gamma}{\cal R}^{\gamma\beta}
=\delta^{\alpha\beta}$ and 
${\cal R}^{z\alpha}=n^{\alpha}$, one can see that $c^{\dagger}
\nv\cdot{\bm \sigma}c= a^{\dagger}\sigma^{z}a$. 
The SU(2) gauge field is given by 
$A_{\mu}= -iU^{\dagger}\partial_{\mu}U 
 \equiv A^{\alpha}_{\mu}\sigma^{\alpha} = {\bm A}_{\mu} 
   \!\cdot\! {\bm \sigma}$ 
$(\mu = 0,x,y,x)$, where $0$ indicates the time component. 
In the rotated frame, the Lagrangian $L$ is given by 
$L = L_{\rm el}-H_{\rm e-A}-\tilde{H}_{\rm so}$ 
up to the first order in $A^{\alpha}_{\mu}$ \cite{com1,com2}, where 
\beqa
\label{L_el}
&&L_{\rm el} = \int d\rv~ a^{\dagger}
\left[i\hbar
\frac{\partial}{\partial t}+
\frac{\hbar^2}{2m}\nabla^{2}
+\vare_{\rm F}
+M \sigma^{z}-V_{\rm imp}
\right]a,
\label{Lagrangian-e}\nonumber\\
&&\\
&&\tilde{H}_{\rm so}= \lambda_{\rm so} \frac{m}{\hbar}
\vare_{ij\alpha}\int d^{3}\rv 
(\partial_{i}V_{\rm imp}(\rv))
{\cal R}^{\alpha\beta}(x)\tilde{j}^{\beta}_{j}(x),\\
&&H_{\rm e-A} = 
\int d\rv~\left[
\tilde{\sigma}^{\alpha}(x)A_{0}^{\alpha}(x)
+
\tilde{\mathcal{J}}^{\alpha}_{i}(x) A^{\alpha}_{i}(x)
\right]. 
\eeqa
 Here $\tilde{\sigma}^{\alpha}=a^{\dagger}\sigma^{\alpha}a$ and 
$\tilde{j}^{\beta}_{j}=\displaystyle\frac{\hbar}{2mi}
(a^{\dagger}\sigma^{\alpha}
\overset{\leftrightarrow}{\partial}_{i}a)$ 
are spin density and spin-current density, respectively, 
in the rotated frame, 
and $\tilde{\cal{J}}^{\alpha}_{i}
= \tilde{j}^{\alpha}_{i} + \tilde{j}^{{\rm so},\alpha}_{i}$
with 
\beqa
 \tilde{j}^{{\rm so},\alpha}_{i} = -\frac{\lambda_{\rm so}}{\hbar} 
\vare_{ij\beta}
(\partial_{j} V_{\rm imp}) {\cal R}^{\alpha\beta} a^{\dagger}a, 
\eeqa 
being an additional spin-current density due to SOI.

{\it Hall conductivity:} 
It is known that dynamics of inhomogeneous magnetization 
produces
a spin motive force, 
and induces a diagonal electric current, as given 
by the first term of Eq. (\ref{total_current}), 
with $\sigma_{\rm s} = \sigma_\uparrow - \sigma_\downarrow$. 
 Here 
$\sigma_{\uparrow \, (\downarrow)} 
= e^{2}n^{\rm el}_{\uparrow \, (\downarrow)}\tau_{\uparrow \, (\downarrow)}/m$ 
is the Drude conductivity for each spin component, 
with 
$n^{\rm el}_{\sigma}$ and 
$1/\tau_{\sigma}= 2\pi n_{\rm imp}u^{2}\nu_{\sigma}/\hbar$ 
($\sigma = \uparrow, \downarrow$) 
being the density and the damping rate, 
respectively, of spin-$\sigma$ electrons. 
($n_{\rm imp}$ is the concentration of impurities). 
In the gauge-field formulation, 
the spin motive field ${\bm E}_{\rm s}$ is given in terms of 
the $z$-component of the SU(2) gauge field $A^{z}_{\mu}$ as 
\cite{{Volovik87},{Tserkovnyak08},{SK08}}
\beqa
\label{SEF}
E_{{\rm s},i} = \frac{\hbar}{e}
(\partial_{i}A^{z}_{0}-\partial_{0}A^{z}_{i}) , 
\label{Es_A}
\eeqa
in precisely the same way as the ordinary electric field is 
given in terms of the electromagnetic vector potential. 
 One can show that this expression (\ref{Es_A}) coincides with 
the expression given in Eq. (\ref{Es}). 
 To study the Hall response to ${\bm E}_{\rm s}$, we here evaluate the Hall 
current as a linear response to the spatial component $A^{z}_{i}$ 
for simplicity. 

The current-density operator, $\hat{\mathcal{J}}_{i}$, 
consists of three parts, 
$\hat{\mathcal{J}}_{i} = {j}_{i} + j^{A}_{i} + j^{\rm so}_{i}$, 
where 
$j_{i} = -\displaystyle\frac{e\hbar}{2mi}
a^{\dagger}\overset{\leftrightarrow}{\partial}_{i}a$, 
$j^{A}_{i} = -\displaystyle\frac{e\hbar}{m}
A^{\alpha}_{i}\tilde{\sigma}^{\alpha}$, 
and 
$j^{\rm so}_{i} = \lambda_{\rm so} \frac{e}{\hbar} 
 \vare_{ij\alpha} (\partial_{j} V_{\rm imp}) 
{\cal R}^{\alpha\beta} (a^\dagger \sigma^{\beta} a)$, 
the last two coming from the local transformation and SOI, respectively. 
 Using Kubo formula, 
the Fourier components of the Hall current density 
are given by 
\beqa
\mathcal{J}_{i}^{\rm H}(\qv,\omega) 
= \sum_{\qv'}\chi_{ij}(\qv,\qv',\omega)
A^{z}_{j}(\qv',\omega), 
\eeqa
where $\chi_{ij}$ $(i\neq j)$ is the correlation function between 
the current and spin-current densities. 
 In the Matsubara frequency representation, they are given by 
\beqa
&& \hskip -12mm 
\chi_{ij}(\qv,\qv',i\omega_{\lambda}) = \chi^{{\rm skew}}_{ij} 
+ \chi^{{\rm side}}_{ij} ,\\
&& \hskip -8mm 
\chi^{{\rm skew}}_{ij}
= -\hbar\int_{0}^{\beta}d\tau e^{i\omega_{\lambda}\tau}
\langle {\rm T}_{\tau} j_{i}(\qv,\tau)\tilde{j}^{z}_{j}(-\qv')\rangle ,
\\
&& \hskip -8mm 
\chi^{{\rm side}}_{ij}
= -\hbar\int_{0}^{\beta}d\tau e^{i\omega_{\lambda}\tau}\nonumber\\
&& \hskip -5mm \times
\langle {\rm T}_{\tau} \{
j_{i}(\qv,\tau)\tilde{j}^{{\rm so},z}_{j}(-\qv')
+j^{\rm so}_{i}(\qv,\tau)\tilde{j}^{z}_{j}(-\qv')\}
\rangle .
\eeqa
 Here $\beta$ is the inverse temperature, 
$\omega_{\lambda}=2\pi \lambda/\beta $
(with $\lambda$ being integer) is the Bosonic Matsubara frequency, 
and $j_{i}(\qv)$, $j^{\rm so}_{i}(\qv)$, 
$\tilde{j}^{z}_{i}(\qv)$ and 
$\tilde{j}^{{\rm so},z}_{i}(\qv)$
are the Fourier components of the current and spin-current densities. 
 The thermal average $\langle \cdots \rangle$ is taken in the equilibrium 
state determined by $L_{\rm el}$ in Eq. (\ref{L_el}). 
 Since the present theory satisfies the Onsager reciprocity relations, 
the following calculation can
be performed in a way similar to 
the spin Hall conductivity \cite{DasSarma06}. 

\begin{figure}[b]
\includegraphics[scale=0.32]{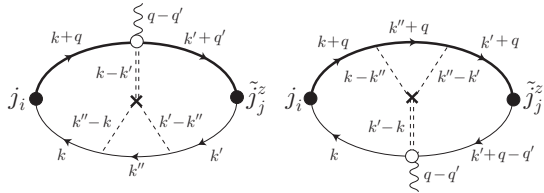}
\caption{
Feynman diagrams for $\chi^{\rm skew}_{ij}$. 
The thick (thin) solid line represents an electron line 
carrying Matsubara frequency $i\vare_{n}+i\omega_{\lambda}$ $(i\vare_{n})$. 
The dotted line (double dotted line with an open circle) represents potential 
(spin-orbit) scattering $V_{\rm imp}$ ($\tilde{H}_{\rm so}$) by impurities. 
 The wavy line represents the rotation matrix ${\cal R}^{\alpha\beta}$. 
}
\label{fig_ss}
\end{figure}

{\it Skew-scattering process:} 
In the lowest order in $\lambda_{\rm so}$, 
the first contribution to $\chi^{\rm skew}_{ij}$ 
comes from the third-order impurity scattering with first order 
coming from $\tilde{H}_{\rm so}$. 
 The diagrammatic expressions are shown in Fig. \ref{fig_ss}. 
 We are interested in slowly varying magnetization 
compared with the characteristic time and length scales of electrons, 
and 
put $\qv,\qv'=0$ in the correlation function $\chi^{\rm skew}_{ij}$ 
related to the electrons. 
After some calculations, we obtain 
\beqa
\label{chi-ss}
&& \hskip -8mm 
\chi^{{\rm skew}}_{ij}(\qv,\qv',i\omega_{\lambda})
= i \lambda_{\rm so} \frac{4e}{9\hbar} n_{\rm imp}^{\phantom{\dagger}} u^{3}
\vare_{ij\alpha} n^{\alpha}_{\qv-\qv'}\nonumber\\
&& \hskip -3mm  \times
\frac{1}{\beta}\sum_{n,\sigma}
\left(\sum_{\kv} \vare_{\kv} G^{+}_{\kv\sigma}G_{\kv\sigma}\right)^{2}
 \sum_{\kv'} (G^{+}_{\kv'\sigma}-G_{\kv'\sigma}),
\eeqa
where $n^{\alpha}_{\qv}$ is the Fourier component of the unit vector 
$n^{\alpha}(\rv)$. 
 The impurity-averaged Green's functions are given by 
$G_{\kv\sigma}(z) 
= (z-\xi_{\kv\sigma}+i\gamma_{\sigma}{\rm sgn}({\rm Im}z))^{-1}$, 
where 
$\xi_{\kv\sigma}= \vare_{\kv} - \vare_{{\rm F}\sigma}$, 
$\vare_{\kv}=\displaystyle\frac{\hbar^{2}\kv^2}{2m}$ 
and $\gamma_{\sigma} = \displaystyle\frac{\hbar}{2\tau_{\sigma}}$, 
and we put  
$G^{+}_{\kv\sigma} = G_{\kv\sigma}(i\vare_{n}+i\omega_{\lambda})$ 
and $G_{\kv\sigma}=G_{\kv\sigma}(i\vare_{n})$. 
 After the analytic continuation, $i\omega_{\lambda} \to \omega + i0$, 
we obtain   
\beqa
\label{chi_ss}
\chi^{{\rm skew}}_{ij}(\qv,\qv',\omega) 
= -i\omega \frac{\hbar}{e}\sigma^{\rm skew}_{\rm SH} 
  \vare_{ij\alpha} n^{\alpha}_{\qv-\qv'}, 
\eeqa
up to ${\cal O}(\omega)$. 
 Here we have put 
$\sigma^{\rm skew}_{\rm SH} 
= \sigma^{\rm skew}_{\uparrow} + \sigma^{\rm skew}_{\downarrow}$ 
with 
\beqa
\sigma^{\rm skew}_{\uparrow \, (\downarrow)} 
= \lambda_{\rm so} u \, \frac{2\pi e^2}{\hbar^{2}}
(n^{\rm el}_{\uparrow \, (\downarrow)})^{2}\tau_{\uparrow \, (\downarrow)}, 
\eeqa
which explicitly depends on the impurity potential $u$ and 
the relaxation time $\tau_{\sigma}$. 

\begin{figure}[b]
\includegraphics[scale=0.3]{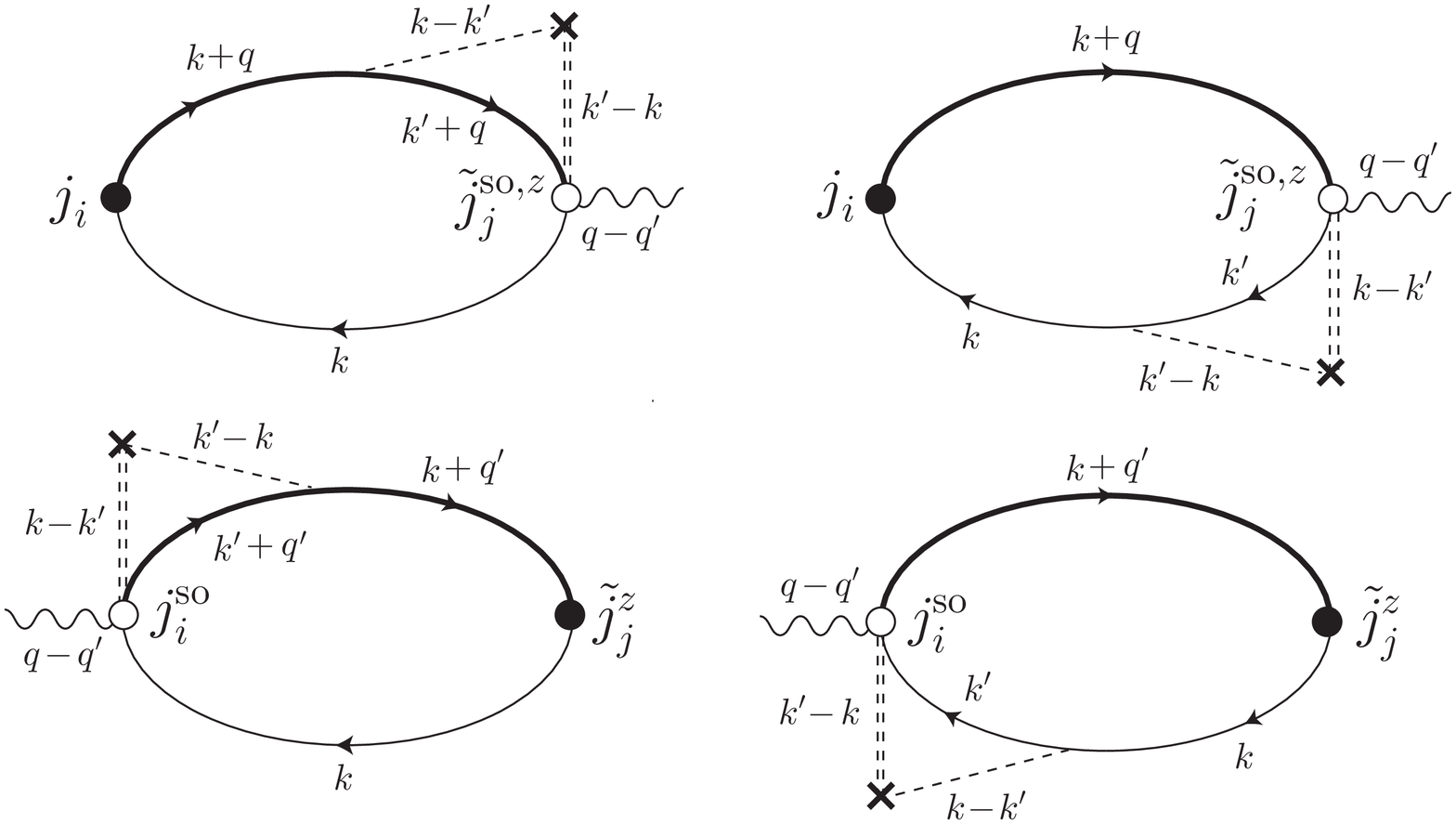}
\caption{Feynman diagrams for $\chi^{\rm side}_{ij}$. 
The meaning of the diagrams is the same as Fig.1. 
} 
\label{fig_sj}
\end{figure}

{\it Side-jump process:} 
In the lowest order in $\lambda_{\rm so}$, 
the first contribution to ${\chi}^{{\rm side}}_{ij}$ 
comes from the second order impurity scattering 
(shown in Fig.~\ref{fig_sj}), 
and is given by 
\beqa
&& \hskip-10mm 
 \chi^{{\rm side}}_{ij}(\qv,\qv',i\omega_{\lambda}) 
= 
  i\lambda_{\rm so} \frac{4e}{3\hbar}n_{\rm imp}^{\phantom{}}u^{2}
\vare_{ij\alpha} n^{\alpha}_{\qv-\qv'}
\nonumber\\
&\times&\frac{1}{\beta}\sum_{n}\sum_{\kv,\kv',\sigma}\vare_{\kv} 
 G^{+}_{\kv\sigma}G_{\kv\sigma}(G^{+}_{\kv'\sigma}-G_{\kv'\sigma}). 
\label{chi-sj}
\eeqa
 After the analytic continuation, $i\omega_{\lambda} \to \omega + i0$, 
we obtain 
\beqa
\label{chi_sj}
 \chi^{{\rm side}}_{ij}(\qv,\qv',\omega) 
= -i\omega \frac{\hbar}{e} \sigma^{\rm side}_{\rm SH} 
\vare_{ij\alpha} n^{\alpha}_{\qv-\qv'}, 
\eeqa
up to ${\cal O}(\omega)$. 
Here 
$\sigma^{\rm side}_{\rm SH} 
= \sigma^{\rm side}_{\uparrow} + \sigma^{\rm side}_{\downarrow}$, 
with 
\beqa
\sigma^{\rm side}_{\uparrow \, (\downarrow)} 
= \lambda_{\rm so}\frac{2e^2}{\hbar} n^{\rm el}_{\uparrow \, (\downarrow)}, 
\eeqa
being independent of the relaxation time. 

Combining Eqs. (\ref{chi_ss}) and (\ref{chi_sj}), 
we obtain the Hall current as 
\beqa
\label{Hall_current1}
\mathcal{J}^{\rm H}_{i}(\rv,t) 
&=& \sigma_{\rm SH}\vare_{i\alpha j} 
n^{\alpha}(\rv,t)
\frac{\hbar}{e}\left(-\frac{\partial}{\partial t}A_{j}^{z}(\rv,t)\right)
\\
&=& \sigma_{\rm SH}(\nv \times {\bm E}_{\rm s})_{i}, 
\label{Hall_current2}
\eeqa
where 
$\sigma_{\rm SH} = \sigma^{\rm skew}_{\rm SH} + \sigma^{\rm side}_{\rm SH}$ 
is the total spin Hall conductivity, and ${\bm E}_{\rm s}$ is given 
by (\ref{SEF}). 
 The Hall current $\mathcal{J}^{\rm H}$ flows in the direction 
perpendicular to both $\nv$ and ${\bm E}_{\rm s}$. 
 This expression is our main result. 
 The total current is given by the sum of the diagonal current 
and the Hall current as Eq.~(\ref{total_current}). 
 Equations (\ref{total_spin_current_E}) and (\ref{total_current_E}) 
can be obtained in a similar manner.

The spin-transfer torque that the $s$-electrons exert on the localized 
$d$-spins is represented by $\sim {\bm j}_{\rm S}\cdot\nabla \nv$
\cite{KTS06}, 
with ${\bm j}_{\rm S} = \sigma_{\rm s} {\bm E}$ 
being the diagonal spin-current density, the first term in 
Eq.(\ref{total_spin_current_E}). 
 The existence of the second term of Eq. (\ref{total_spin_current_E})
suggests the existence of a spin-transfer torque 
due to the spin Hall current, and 
our result (\ref{Hall_current2}) of the spin Hall motive force 
should be the reaction to this torque. 
 Such a study will be reported elsewhere. 

 The (spin) Hall resistivity is given by 
$\rho_{\rm SH} = \sigma_{\rm SH}/\sigma^2_{\rm s}
\sim P^{-3}\sigma_{\rm H}/\sigma^2_{\rm c}$, 
where 
$\sigma_{\rm H} = \sigma_{{\rm H} \uparrow} - \sigma_{{\rm H} \downarrow}$,  
with 
$ \sigma_{{\rm H} \uparrow \, (\downarrow)} 
 = \sigma^{\rm skew}_{\uparrow \, (\downarrow)} 
 + \sigma^{\rm side}_{\uparrow \, (\downarrow)} $, 
is known as the extrinsic anomalous Hall conductivity \cite{DCB01}. 
 For a typical value of $P\sim 0.5$, $\rho_{\rm SH}$ in the present case 
is one order of magnitudes larger than that of the conventional AHE.

\begin{figure}[h]
\includegraphics[scale=0.22]{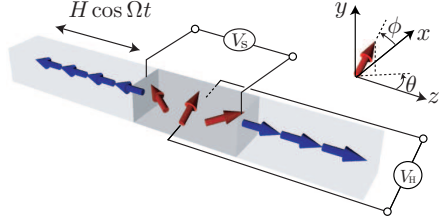}
\caption{Schematic illustration of an experimental setup for the detection 
of the inverse spin Hall motive force caused by a field-driven 
domain wall oscillation.}
\label{fig_AHE}
\end{figure}

{\it DW oscillation:} 
 Let us apply the result (\ref{Hall_current2}) to a magnetic field driven 
domain wall (DW) oscillation in a ferromagnetic nanowire. 
 We consider a Hall device, as shown in Fig. \ref{fig_AHE}, 
where the cross section of the wire forms a square, 
which allows us to neglect hard axis anisotropy energy, 
and the one-dimensional tail-to-tail DW is positioned at $z=0$. 
 When an ac magnetic field is applied along the wire $(// \hat z)$, 
spins in the wall oscillate around the $z$ axis. 
 Taking the ac field as $H_{\rm ac} (t)=H\cos \Omega t$, 
where $H$ is the amplitude and $\Omega$ is the frequency, 
and solving the Landau-Lifshitz-Gilbert equation for $\nv$ analytically, 
we obtain a DW solution
$\nv =\left(\frac{\cos\phi}{\cosh\frac{z-X}{\lambda}},
\frac{\sin\phi}{\cosh\frac{z-X}{\lambda}},
\tanh\frac{z-X}{\lambda}\right)$, 
where $\phi=\frac{1}{1+\alpha^{2}}(\gamma H/\Omega)\sin\Omega t$ 
and $X=\alpha \lambda \phi$.  
 Here $\gamma$ is the gyromagnetic constant, 
$\lambda$ is the width of the DW, 
and $\alpha$ is the Gilbert damping constant. 
 Substituting this solution into Eq. (\ref{SEF}), 
we obtain the spin motive force as 
\beqa
V_{\rm s}=\int_{-\infty}^{\infty}dz
E_{{\rm s},z}=-\frac{\hbar}{e}
\frac{\gamma H}{1+\alpha^{2}}
\cos\Omega t,  
\eeqa
which oscillates in time. 
 For an open circuit condition in the lateral face of the wire, 
$\mathcal{J}_{x}=\mathcal{J}_{y}=0$, 
the Hall voltage at the DW center is obtained as 
\beqa
V_{\rm H} = \frac{\sigma_{\rm SH}}{\sigma_{\rm s}}\frac{w}{2\lambda}
V_{\rm s}, 
\eeqa
where $w$ is the width of the wire. 
 If we choose $w\simeq\lambda$, $\alpha=0.01$, 
$\gamma H =\Omega = 100~{\rm MHz}$, and 
$\sigma_{\rm SH}/\sigma_{\rm s}\simeq P^{-2}\sigma_{\rm AH}/\sigma_{\rm c}
\sim 1$, 
the amplitude of $V_{\rm H}$ is estimated as $|V_{\rm H}|\sim 31~{\rm nV}$, 
which might be detectable experimentally.  
  For a head-to-head DW, the phase of $V_{\rm s}$ and $V_{\rm H}$ changes 
by $\pi$ relative to $H_{\rm ac} (t)$, and this fact may be used to 
discriminate the true signal.

A dc magnetic field applied in the same (easy-axis) direction 
can also lead to an oscillatory dynamics by the Walker's breakdown 
\cite{Schryer74}, and this will produce ac signals $V_{\rm s}$ 
and $V_{\rm H}$ similar to the ones obtained above.


In conclusion, we have presented a microscopic theory of 
the AHE driven by the spin motive force due to inhomogeneous spin dynamics. 
 It is shown that a Hall current is induced by the spin motive force 
in the presence of (extrinsic) spin-orbit interaction, and 
the corresponding Hall resistance is enhanced compared with the 
conventional AHE. 
 Applying the result to the field driven domain-wall oscillation, 
we have shown that a Hall voltage is generated 
in the lateral face of the wire.

The authors would like to thank G. Tatara, Y. Nakatani and E. Saitoh 
for valuable discussions. 
We also thank Q. Niu and S. A. Yang for sending Ref.\,\cite{YBKXNTE08} to us 
before publication. 
This work is partially supported by a Grant-in-Aid from Monka-sho, Japan. 
J. S. thanks The Kurata Memorial Hitachi Science and Technology Foundation 
and The Sumitomo Foundation 
for financial support.


\begin{thebibliography}{99}
\bibitem{spin_dependence}
The spin dependence of ${\bm E}_{\rm s}$ 
can be made explicit by multiplying the
right-hand side of Eq. (\ref{Es}) by a factor ${\bm n} \cdot {\bm \sigma}$
which is $+1$ ($-1$) for majority- (minority-) spin electrons.
\bibitem{com1}
Actually, because of the identity, 
$\partial_\mu {\bm A}_\nu - \partial_\nu {\bm A}_\mu 
 = 2 {\bm A}_\mu \times {\bm A}_\nu$, 
we need to retain terms up to the second order in ${\bm A}_\mu$. 
 We have, however, observed in the calculation that the second-order 
terms cancel out and only the first-order terms survive 
in the final expression through ${\bm E}_{\rm s}$, Eq.(\ref{Es_A}).
\bibitem{com2}
The expansion parameter of the present gauge-field treatment is 
$\ell /{\lambda_{\rm w}}$ 
and 
$\omega_{\rm w}\tau_{\sigma}$ for $A^z_\mu$, 
and 
$\displaystyle
\frac{\vare_{\rm F}}{M}\frac{1}{k_{\rm F}
\lambda_{\rm w}}\frac{\ell}{\lambda_{\rm w}}$ 
and 
$\displaystyle \frac{\hbar \omega_{\rm w}^2 \tau_{\sigma}}{M}$ 
for $A^x_\mu$ and $A^y_\mu$ \cite{TKS08} . 
($k_{\rm F}$ and $l$ are the Fermi wavenumber and the mean free path 
of electrons, and $\lambda_{\rm w}$ and $\omega_{\rm w}$ are 
characteristic length and frequency scales of the magnetization.)
\end{thebibliography}
\end{document}